\documentstyle[12pt]{article}
\def\L{{\cal L}}
\def\ka{{\kappa}}
\def\la{{\lambda}}

\def\ga{{\gamma}}
\newcommand{\be}{\begin{eqnarray}}
\newcommand{\en}{\end{eqnarray}}
\newcommand{\nn}{\nonumber}
\begin{document}
\begin{titlepage}
\setlength{\textwidth}{5.9in}
\begin{flushright}
EFI 99-01\\
MPI-Ph/99-04  \\
\end{flushright}
\begin{center}
\vskip 0.3truein
{\Large\bf {Reduction of Coupling Parameters}} \\
\vskip 0.15truein
{\Large\bf {and Duality}} 
\footnote{Dedicated to Wolfhart Zimmermann on the Occasion of his 70th Birthday. \\
To appear in {\it Recent Developments in Quantum Field Theory}, 
Springer Verlag, Heidelberg, New York. (Editors: P. Breitenlohner, D. Maison and J. Wess).
Ringberg Symposium, June 1998.}
\vskip0.4truein
{Reinhard Oehme}
\footnote{E-mail: oehme@theory.uchicago.edu}
\vskip0.4truein
{\it Enrico Fermi Institute and Department of Physics}\\
{\it University of Chicago} \\
{\it Chicago, Illinois, 60637, USA}
\footnote{Permanent Address}\\
{\it and}\\
{\it Max-Planck-Institut f\"{u}r Physik}\\
{\it - Werner-Heisenberg-Institut -}\\
{\it 80805 Munich, Germany}
\end{center}
\vskip0.2truein
\centerline{\bf Abstract}
\vskip0.3truein
The general method of the reduction in the number of coupling
parameters is discussed. Using renormalization group invariance, 
theories with several independent couplings are related to a set
of theories with a single coupling parameter. The reduced theories 
may have particular symmetries, or they may not be related
to any known symmetry. The  method is more general than the 
imposition of invariance properties. Usually, there are 
only a few reduced theories with an asymptotic power series expansion 
corresponding to a renormalizable Lagrangian. There also exist
`general' solutions containing non-integer powers and 
sometimes logarithmic factors.
As an example for the use of the reduction method, the dual magnetic 
theories associated with certain supersymmetric gauge theories
are discussed. They have a superpotential with a Yukawa coupling
parameter. This parameter is expressed as a function of the gauge
coupling. Given some standard conditions, a unique,
isolated power series solution of the reduction equations is 
obtained. After reparametrization, the Yukawa coupling is 
proportional to the square of the gauge coupling parameter. The
coefficient is given explicitly in terms of the numbers of colors and
flavors. `General' solutions with non-integer powers are also
discussed. 
A brief list is given of other applications of the reduction method.  
\end{titlepage}
\newpage
\baselineskip 18 pt
\pagestyle{plain}
\setlength{\textwidth}{5.9in}
\setcounter{equation}{0}

\vskip0.2truein

{\bf 1. Introduction.}

The method of reduction in the number of coupling
parameters \cite{OZR,OZS,WZR,OSZ,RON},
\cite{CON,WZK,KSP,KMO,ROR}, \cite{CHE} has 
found many theoretical and phenomenological
applications. It is a very general method, based essentially
upon the requirement of renormalization group invariance
of the original multi-parameter theory, as well as the related
reduced theories with fewer couplings. Combining the
renormalization group equations of original and reduced
theories, we obtain a set of {\it reduction equations}. These
are differential equations for the removed couplings considered
as functions of the remaining parameters. They are necessary
and sufficient for the independence of the reduced theories
from the normalization mass. We consider massless theories, or
mass independent renormalization schemes, so that no mass 
parameters appear in the coefficient functions of the renormalization
group equations. This can be arranged, provided the original
coefficient functions have a well defined zero-mass limit
\cite{WZM}.

In this paper, we discuss only reductions to a single coupling,
which covers most cases of interest. Usually, we can choose one
of the original couplings as the remaining parameter. The multi-
parameter theory is assumed to be renormalizable with an asymptotic
power series expansion in the weak coupling limit. However the
reduced theories, as obtained from the reduction equations, may
well not all have such expansions in the remaining coupling. Non-
integer powers and logarithms can appear, often with undetermined
coefficients. Such {\it general solutions} do not correspond to
conventional renomalized power series expansions associated with
a Lagrangian. But they are still well defined in view of their
embedding in the renormalized multi-parameter theory. Nevertheless,
it is the relatively small number of uniquely determined power series
solutions of the reduction equations, which is of primary interest.
Depending upon the character of the system considered, there may
be additional requirements which further reduce the number of these
solutions. Although we consider renormalizable theories, with
appropriate assumptions, the reduction method can also be applied 
in cases where the original theory is non-renormalizable.

Regular reparametrization is a very useful tool in connection with
the reduction method. For theories with two or more coupling
parameters, it is not possible to reduce the $\beta$-function
expansions to polynomials. However, in the reductions to one coupling,
we can usually remove all but the first term in the power series
solutions of the reduction equation with determined coefficients.
The $\beta$-functions of the corresponding reduced theories remain
however infinite series. As seen from many examples, these
reparametrizations lead to frames which are very natural for the
reduced theories.

The imposition of a symmetry on the multi-parameter theory is
a conventional way of relating the coupling parameters. If there
appear no anomalies, we get a renormalizable theory with fewer
parameters so as to implement the symmetry. These situations are
all included in the reduction scheme, but our method is more general,
leading also to unique power series solutions which exhibit no
particular symmetry. This situation is illustrated by an example we
have included. An $SU(2)$ gauge theory with matter fields in the
adjoint representation. Besides the gauge coupling, there are three
additional couplings. With only the gauge coupling remaining after
the reduction, we get two acceptable power solutions. One of the
reduced theories is an $N=2$ supersymmetric gauge theory, while
the other solution leads to a theory with no particular symmetry.

The main example presented in this article is connected with 
{\it duality} \cite{SEI,WIS,SEN,APS}. We consider $N=1$ supersymmetric QCD (SQCD) 
and the corresponding dual theory, magnetic SQCD. 
The primary interest is in the phase
structure of the physical system described by these theories.
Essential aspects of this phase structure were first 
obtained on the basis of supercovergence
relations and BRST methods \cite{LOP,ROC,RLP}, and more recently with the help of duality.
\cite{SEI,SEN}.  
We exhibit the quantitative agreement of both approaches \cite{NVS,DSP}. While
duality is formulated only in connection with supersymmetry, the 
supercovergence arguments can be used also for QCD and similar
theories \cite{ROC,RLP}, \cite{NIC}. Of particular interest is the transition point at 
$N_F = \frac{3}{2}N_C$ for SQCD \cite{LOP,SEN}, where $N_F$ and $N_C$ are the numbers
of flavors and colors respectively. It is the lower end of the conformal
window. For smaller values of $N_F$, the quanta of free, electric SQCD 
are confined, the system is described by free magnetic excitations
of the dual theory (for $N_C>4$), and eventually by mesons and baryons.
(The corresponding transition point for QCD is given 
by $N_F = \frac{13}{4}N_C$ ).  

As the original theory, SQCD has only 
the gauge coupling $g_e$. The dual theory is constructed on the basis
of the anomaly matching conditions \cite{SEI,SEN}. It
involves the two coupling parameters $g_m$ and $\la_1$, where $\la_1$ is a Yukawa
coupling associated with a superpotential. This potential is required by
duality, mainly since theories, which are dual to each other, 
must have the same global symmetries.

At first, we apply the reduction method to the magnetic theory in the
conformal window $\frac{2}{3}N_C <N_F < 3N_C$ \cite{DSP,RDT}. We find
two power series solutions. After reparametrization, one solution is given
by $\la_1(g_m^2) = g_m^2 f(N_C,N_F)$, with $f$ being a known function
of the numbers of colors and flavors for SQCD. The other solution is
$\la_1(g_m^2) \equiv 0$. Since the latter removes the superpotential, it
is excluded, and we are left with a unique single power solution.
This solution implies a theory with a single gauge coupling $g_m$,
and renormalized perturbation expansions which are power series in 
$g_m^2$. It is the appropriate dual of SQCD. There are `general' solutions,
but they all approach the excluded power solution $\la_1(g^2) \equiv 0$.
With one exception, they involve non-integer powers of $g_m^2$.
The reduction can be extended to the `free electric region' $N_F > 3N_C$,
and to the `free magnetic region' $N_C+2 < N_F < \frac{2}{3}N_C$, $(N_C > 4$).
The results are similar, and discussed in detail in \cite{RDT}. 
In the free magnetic case, we deal however with the approach to a trivial
infrared fixed-point. 

\noindent Possible connections of the reduction results with features of
brane dynamics remain to be considered. Internal fluctuations of branes
may be of relevance for the field theory properties obtained here. 
 
\vskip0.2truein

{\bf 2. Reduction Equations}

We consider renormalizable quantum field theories with several
coupling parameters. It is assumed that there is a mass-independent
renormalization scheme, so that no mass parameters occur in the 
coefficient functions of the renormalization group equations. Let
$\la,\la_1,\ldots,\la_n$ be $n+1$ dimensionless coupling parameters
of the theory. One can reduce this system in various ways, but we want
to consider the parameter $\la$ as the primary coupling, and express
the remaining $n$ couplings as functions of  $\la$:
\be
\la_k = \la_k (\la), ~~~k = 1,\ldots,n ~.
\en 
It is assumed, that these functions $ \la_k (\la)$ are independent of the 
renormalization mass $\ka$, which can always be arranged. 

The Green's functions  $G \left( k_i, \ka^2, \la, \la_1,\ldots,
\la_n\right)$ of the original multi-parameter version of the theory
satisfy the usual renormalization group equations with the coefficient
functions $\beta, \beta_{k}$, and the anomalous dimension $\gamma_{G}$,
which depend upon the $n+1$ coupling parameters. The corresponding
Green's functions of the reduced theory are given by
\be
G(k_i,\ka^2,\la) = G \left( k_i, \ka^2, \la, \la_1 (\la),\ldots,
\la_n (\la)\right).
\en
Renormalization group invariance requires that they satisfy the equation
\be
\left( \ka^2 \frac{\partial}{\partial \ka^2} + \beta(\la)
\frac{\partial}{\partial\la} + \gamma_G(\la) \right)
G(k_i,\ka_2,\la) ~ = ~ 0 ~~,
\en 
where $\beta(\la)$ and $\ga_{G}(\la)$ are given by the corresponding 
original coefficients with the insertions $\la_k = \la_k (\la), ~~~k = 1,\ldots,n ~$.
Comparison of Eq.(3) with the original multi-parameter renormalization group
equation implies then
\be
\beta(\la) \frac{d\lambda_k (\la)}{d\la} = \beta_k (\la)~,
~~~k=1,\ldots,n  
\en
These are the {\it Reduction Equations}, which are necessary and sufficient
for the validity of Eq.(3).

It is of interest to briefly consider the relationship between the reduction method
as described above, and the equations for the effective coupling functions
$\overline{\lambda}(u),\overline{\lambda}_k(u)$,
where $u$ is
the dimensionless scaling parameter $u=k^2/\kappa^2$. These functions satisfy the
equations
\be
u\frac{d \overline{\lambda}}{d u}
&=&
\beta(\overline{\lambda},\overline{\lambda}_1,
\dots,\overline{\lambda}_n)~,\nn\\
u\frac{d \overline{\lambda}_k}{d u}
&=&\beta_k(\overline{\lambda},\overline{\lambda}_1,
\dots,\overline{\lambda}_n)~.
\en
With $\overline{\lambda}(u)$ being an analytic function, we can choose a point
where $(d \overline{\lambda}(u)/du)\neq 0$ and introduce $\overline{\lambda}(u)$
as a new variable Eqs.(5,6). The result is again the reduction equations (4).

With effective couplings, we study the multi-parameter theory at different mass scales.
In the reduction method, we consider the set of different field theories with one
coupling parameter (or a reduced number), which can be obtained from a given
multi-parameter theory as solutions of the reduction equations. The elements of
this set are labeled by the free parameters of the solution, and all are considered
at the same fixed mass scale. With some natural assumptions the number of theories
in this set is usually smaller than the number of original coupling parameters,
and the different theories have characteristic physical and mathematical features.
This is best seen in examples, some of which we discuss below. It must be remembered,
that the origin of the coupling parameter space is a singular point, so that the
Picard-Lindeloef theorem about the uniqueness of solutions at regular points does
not apply.
 
As described so far, the reduction scheme is very general, but in practice we usually
know the $\beta$-functions only as asymptotic expansions in the small coupling limit.
Within the framework of renormalized perturbation theory, we restrict ourselves here
to expansions of the form
\be
\beta(\lambda,\lambda_1,\dots,\lambda_n)
&=&\beta_{0}\lambda^2+
(~\beta_{1}\lambda^3+
\beta_{1k}\lambda_k\lambda^2+
\beta_{1k k'}\lambda_k\lambda_{k'}\lambda~)\nn\\
& &+\sum_{n=4}^{\infty}\sum_{m=0}^{n-1}
\beta_{n-2,k_1,\dots,k_m}\lambda_{k_1}\cdots
\lambda_{k_m}\lambda^{n-m}~,\nn\\
\beta_{k}(\lambda,\lambda_1,\dots,\lambda_n)
&=&(c_{k}^{(0)}\lambda^2+
c_{k, k'}^{(0)}\lambda_{k'}\lambda+
c_{k, k' k''}^{(0)}\lambda_{k'}\lambda_{k''}~)~,\nn\\
& &+\sum_{n=3}^{\infty}\sum_{m=0}^{n}
c_{k,k_1, \dots,k_m}^{(n-2)}\lambda_{k_1}\cdots
\lambda_{k_m}\lambda^{n-m}~.
\en  
In writing the expansions (6), we have assumed that the primary coupling
$\la$ is chosen such that $\beta(0,\la_1,...,\la_n)~=~0$.

With the original $\beta$-functions given as asymptotic power series 
expansions, we will consider in the following solutions $\la_k(\la)$ of the reduction
equations, which are also of the form of asymptotic expansions. Of special interest
are solutions which are power series expansions. But in general, non-integer
powers as well as logarithmic terms are possible.

\vskip0.2truein 

{\bf 3. Power Series Solutions}

Let us first consider solutions of the reduction equations (4) which are asymptotic 
power series expansions. Then the Green's functions $G(k_i,\kappa^2;\la)$ of the
reduced theory have power series expansions in $\la$ and are associated with a 
corresponding renormalizable Lagrangian. It is reasonable to write
\be
\la_k (\la) = \la f_k (\la),~~~ k=1,\ldots,n ~,
\en
where the functions $f_k (\la)$ are bounded for $\la\rightarrow0$ so that $\la_k(0)=0$.
According to the reduction equations, if we had $\la_k(0)\neq0$, the vanishing
of $\beta(0,\la_1(0),...,\la_n(0))$ would imply that also $\beta_k(0,\la_1(0),...,\la_n(0))$
vanishes, which
is too strong a restriction and not fulfilled by Eq.(6). In terms of the functions
$f_k (\la)$ the reduction equations are of the form
\be
\beta \left( \la \frac{d f_k}{d\la} + f_k\right) = \beta_k ~,
\en
where we have introduced the $\beta$-functions
\be
\beta(\lambda)&=&
\beta(\lambda,\lambda f_1,\dots,\lambda f_n)
=\sum_{n=0}^{\infty}~\beta_n(f) \lambda^{n+2}~,\\
\beta_k(\lambda)&=&
\beta_k(\lambda,\lambda f_1,\dots,\lambda f_n)
=\sum_{n=0}^{\infty}~\beta_k^{(n)}(f) \lambda^{n+2}~.
\en
Here the argument $f$ stands for
$f_1(\lambda),\dots,f_n(\lambda)$.
The coefficients are easily obtained from Eqs.(6).
For example, the one-loop terms are given by
\be
\beta_0(f)
=\beta_0,~~~~
\beta_{k}^{(0)}(f)
=c_{k}^{(0)}+c_{kk'}^{(0)}f_{k'}+
c_{kk'k''}^{(0)}f_{k'} f_{k''}~.
\en
For the functions $f_k(\la)$, we write the expansions
\be
f_k(\lambda)=f_{k}^{0}+\sum_{m=1}^{\infty}\chi_{k}^{(m)}\lambda^m ~,
\en
and insert them, together with the series (9) and (10), into the reduction
equations. At the one-loop level, there result then the relations
\be
\beta_k(f^0)-f_{k}^{0}\beta_0)&=&0~,
\en
or in explicit form using Eq.(11),
\be
c_{k}^{(0)}+
(c_{kk'}^{(0)}-\beta_0\delta_{k k'})f_{k'}^{0}+
c_{kk'k''}^{(0)}f_{k'}^{0}f_{k''}^{0}=0~.
\en
These are the fundamental formulae for the reduction.  

Given a solution $f_k^0$ of the quadratic equations (14), we obtain for the expansion
coefficients $\chi_{k}^{(m)}$ the relations
\be
\left(M_{kk^\prime} (f^0) - m \beta_{0} \delta_{kk^\prime}\right)
\chi^{(m)}_{k^\prime} = \left(\beta_{m} (f^0) f^0_k -
\beta^{(m)}_k (f^0)\right) + X^{(m)}_k ~,  
\en 
where $m=1,2,\ldots,~~~ k=1,\ldots, n  ~$.
The matrix $M(f^0)$ is given by
\be
M_{kk^\prime} (f^0) = c^{(0)}_{k,k^\prime} + 2 c^{(0)}_{k,k^\prime
k^{\prime\prime}} f^0_{k^{\prime\prime}} - \delta_{kk^\prime} 
\beta_{0}~.
\en
The rest term $X^{(m)}$ depends only upon the coefficients
$\chi^{(1)},\ldots,\chi^{(m-1)}$, and upon the $\beta$--function
coefficients in (9) and (10), evaluated at
$f_k = f^0_k$, for order $m-1$ and lower.  They vanish for
$\chi^{(1)} = \ldots = \chi^{(m-1)} = 0$.

We see that the {\em one--loop} criteria
\be
det \left( M_{kk^\prime}(f^0)- m \beta_{0}\delta_{kk^\prime}\right)
 \not = 0
   ~~ for ~~m=1,2,\ldots
\en
are sufficient to insure that all coefficients $\chi^{(m)}$ in the
expansion (12) are determined.  Then the reduced theory has a
renormalized power series expansion in $\la$.  All possible solutions
of this kind are determined by the one--loop equation (13) for
$f^0_k$.

With the coefficients $\chi^{(m)}$ fixed, we can use regular {\it reparametrization 
transformations} in order to remove all but the first term in the expansion (12) of the
functions $f_k(\la)$. These reparametrization transformations are of the form
\be
\la^\prime &=& \la^\prime (\la,\la_1,\ldots,\la_n) = \la + a^{(20)}
\la^2 + a_k^{(11)} \la_k \la + \cdots  ~~, \cr
\la^\prime_k  &=&  \la^\prime_k
(\la,\la_1,\ldots,\la_n) = \la_k +
b^{(20)}_{kk^\prime k^{\prime\prime}} \la_{k^\prime}
\la_{k^{\prime\prime}}
+ b^{(11)}_{kk^\prime} \la_{k^\prime} \la + \cdots ~.
\en
They leave invariant the one-loop quantities
\be
f^0_k,~ \beta_{0} (f^0),~ \beta_k^{(0)} (f^0),~
M_{kk^\prime} (f^0) ~. 
\en   
Given the condition (17), we then have a frame where 
\be
 \la_k (\la) = \la f^0_k ~ .
\label{20}
\en
This result is valid to all orders of the asymptotic expansion and determined
by one-loop information. With the expressions (20), the $\beta$-function
expansions (9) and (10) of the reduced theory have constant coefficients 
$\beta_m(f^0),~\beta_k^{(m)}(f^0)$, but they are generally not polynomials. They
satisfy the relations
\be
\beta^{(m)}_k(f^0)~-~f^0_k\beta_m(f^0)~=~0.
\en
for all values of $m$. Only the relations for $m=0$ are reparametrization
invariant. They are the fundamental formulae (13).   
 
So far, we have implicitly assumed that $f^0_k\neq0$. But it is straightforward
to include the cases where  $f^0_k=0$. They are of particular interest for
supersymmetric theories. Suppose we have a solution of the reduction equations
with the asymptotic expansion   
\be
f_k(\lambda) = \chi_{k}^{(N)}\lambda^N+\sum_{m=N+1}^{\infty}\chi_{k}^{(m)}\lambda^m ~,
\en
where  $N\ge 1$ and $\chi_{k}^{(N)}\not= 1$. Then coefficients appearing in this 
equation are again determined except
for the first one, which is invariant. Hence, using regular reparametrization,
there is a frame where
\be
f_k(\lambda) = \chi_{k}^{(N)}\lambda^N .
\en

We have considered here only expansions at the origin in the space of coupling
parameters. However, one can use the method also in connection with
any non-trivial fixed point of the theory.

\vskip0.2truein 

{\bf 2. General Solutions}

At first, let us briefly consider the case where the determinant appearing in Eq.(17)
vanishes. Suppose there is a positive eigenvalue of the matrix $\beta_0^{-1}M(f^0)$
for some $m=N\le1$, $\beta_0\ne0$. Then the asymptotic power series must be supplemented
by terms of the form $\la^m(lg\la)^p$, with $m \le N$ and $1<p<\sigma(N)$. After 
reparametrization, we obtain then an expansion of the form
\be
f_k(\la ) = f^0_k + \chi^{(N,1)}_k \la^N \lg \la +
 \chi_k^{(N)} \la^N + \ldots ~~,
\label{3.12}
\en
All parameters in Eq.(24) are determined except the vector $\chi_k^{(N)}$,
which contains as many free parameters as the degeneracy of the eigenvalue.
Even though the theory considered here can have logarithmic terms in the 
asymptotic expansion, it is `renormalized' in view of it's embedding into
the original, renormalized multi-parameter theory. In special cases it may happen
that the coefficients of the logarithmic terms vanish, as in the example of the
massless Wess-Zumino model.

We now return to systems with non-vanishing determinant for all values of $m$. In addition
to the power series solutions described before, there can be {\it general} solutions of the
reduction equations, which approach the latter asymptotically. In order to describe a 
characteristic case, we assume that $\beta_0\ne0$ and that the matrix $\beta_0^{-1}M(f^0)$
has one positive eigenvalue $\eta$ which is non-integer, with all others being negative.
Then the reduction equations (4) have solutions of the form
\be
f_k (\la) = f^0_k + \sum_{a,b} \chi_k^{(a\eta + b)} \la^{a\eta +
b} + \sum_m \chi^{(m)}_k \la^m
\label{4.1}
\en
with $a=1,2,\ldots~$, $b=0,1,\ldots ~$, $a\eta + b = ~$non-integer. 
After reparametrization, powers with $m<\eta$ are removed, and we have 
\be
f_k(\la) = f^0_k + \chi^{(\eta)}_k \la^{\eta} + \ldots .
\en  
In this expansion all coefficients are determined except $\chi^{(\eta)}_k$, which
may contain up to $r$ arbitrary parameters if the eigenvalue $\eta$ is r-fold
degenerate:
\be
\chi^{(\eta)}_k = C_1 \xi^{(1)}_k + \ldots + C_r \xi^{(r)}_k ~,
\label{4.1b}
\en
where the $\xi^{(i)}_k$ are the eigenvectors.

The results described above can be generalized to situations with several positive,
non-integer eigenvalues. In special cases, where the matrix also has a zero eigenvalue,
logarithmic factors may appear.

So far, we have assumed that $\beta_0\ne0$, and obtained general solutions
which approach the power series solution (20) asymptotically with a power law
as indicated in Eq.(26). The situation is quite different if $\beta_0=0$. 
Then th Matrix $M$ is given by 
\be
M_{kk'}(f^0)=\left(\frac{\partial\beta^{(0)}_k(f)}{\partial{f_{k'}}}\right)_0
\en
and we find that the general solutions and the power series solutions differ
asymptotically by terms which vanish exponentially. We refer to \cite{OZS}
for more details.

Besides the general solutions, which approach the power series solutions
asymptotically, there can be others which move away in the limit
$\la\rightarrow0$. These are not calculable unless the $\beta$-functions are known
more explicitly. However, we can get information about the existence or non-existence
of such solutions on the basis of the linear part of the reduction differential
equations (4). We find that the theorems of Lyapunov \cite{LJA} , with generalizations
by Malkin \cite{MAL}, are applicable here \cite{OSW}. We refer to \cite{RON} 
for some more discussion, and to \cite{KRS} for an application. Generally, it turns out 
that a power series solution (20) is asymptotically stable if there are no
negative eigenvalues of the matrix $\beta_{0}^{-1}M(f^0)$ (or the matrix
$\beta_{N}^{-1}M(f^0)$ in the case of the solution (23)). A solution is unstable if there
is at least one negative eigenvalue.

\vskip0.2truein 

{\bf 4. Gauge Theory}

It should be most helpful to discuss briefly an example. We use a gauge theory
with one Dirac field, one scalar and one pseudoscalar field, all in the adjoint
representation of SU(2) \cite{OSZ}. 
Besides the usual gauge couplings, the direct interaction
part of the Lagrangian is given by
\begin{eqnarray}
{\L}_{dir.int.} &=& i\sqrt{\la_1}~
\epsilon^{abc} \overline{\psi}^a (A^b + i\gamma_5 B^b ) \psi^c \cr 
               &-& \frac14 \la_2 (A^a A^a + B^a B^a)^2 +
\frac14 \la_3 (A^a A^b + B^a B^b)^2 ~~.
\end{eqnarray} 
Writing $\la=g^2$, where $g$ is the gauge coupling, and $\la_k=\la f_k$, with k=1,2,3 ,
the one-loop $\beta$-function coefficients of this theory are given by
\be
(16\pi^2)\beta_{g0} &=& -4 \nn\\
(16\pi^2)\beta_1^{0} &=& 8f_1^2-12f_1 \nn\\
(16\pi^2)\beta_2^{0} &=& 3f_3^2-12f_3f_2+14f_2^2+8f_1f_2-8f_1^2-12f_2+3 \nn\\
(16\pi^2)\beta_3^{0} &=& -9f_3^2+12f_3f_2+8f_3f_1-12f_3-3 .
\en
The algebraic reduction equations (14) have four real solutions, which are given
by
\be 
f^0_1&=&1,~~~~f^0_2=1,~~~~~~f^0_3=1 \nn\\
f^0_1&=&1,~~~~f^0_2=\frac{9}{\sqrt{105}},~~f^0_3=\frac{7}{\sqrt{105}},
\en
and two others with reversed signs of $f^0_2$ and $f^0_3$, so that the classical
potential approaches negative infinity with increasing magnitude of the scalar
fields. These latter solutions will not be considered further. We note that the 
Yukawa coupling is required for the consistency of the reduction.

The eigenvalues of the matrix $\beta_{g0}^{-1}M(f^0)$ are respectively
\be
\left(-2, -3, +\frac{1}{2}\right)
\en
and
\be
\left(-2, -\frac{3}{4}~\frac{25+\sqrt{343}}{\sqrt{105}}, -\frac{3}{4}~\frac{25-\sqrt{343}}{\sqrt{105}}\right)
=(-2, -3.189...,  -0.470...).
\en
There are no positive integers appearing in the equations (32) or (33). Hence the coefficients
of the power series solutions are determined and can be removed by reparametrization,
except for the invariant first term.
With $\la = g^2$ as the primary coupling, $g$ being the gauge coupling, these solutions are
\be
(a) ~~~\la_1 = \la_2 = \la_3 = g^2 ~, 
\en
which corresponds to an $N = 2 $ extended SUSY Yang-Mills theory, and
\be
(b) ~~~\la_1 = g^2,~~ \la_2 = \frac{9}{\sqrt{105}}~ g^2, ~~
\la_3 = \frac{7}{\sqrt{105}}~  g^2 ~,
\en
which is not associated with any known symmetry, at least in four dimensions.
Both theories are `minimally' coupled gauge theories with matter fields. The eigenvalues 
of the matrix $\beta_{g0}^{-1}M(f^0)$, given in Eqs.(32),(33), are all negative with the
exception of the third one for the N=2 supersymmetric theory. In this case we have a general
solution corresponding to Eq.(26) with $\eta=+\frac{1}{2}$, and  with the coefficient given
by $\chi^{(\frac{1}{2})}=(0,C,3C)$, where $C$ is an arbitrary parameter. The theory with
$C\ne0$ corresponds to one with hard breaking of SUSY. It has an asymptotic power series in
$g$ and not in $g^2$, as is the case for the invariant theory.

As we see from Eqs.(32) and (33), both power series 
solutions have some negative eigenvalues of
the matrix $\beta_{g0}^{-1}M(f^0)$, and are therefore unstable. Not all nearby solutions 
approach them asymptotically.

From the present example, and many others, we realize that the special frame, where
the power series solutions of the reduction equations are of the simple form (20), is 
a natural frame as far as the reduced one-parameter theories are concerned. The $\beta$-
functions of the reduced theories are still power series and are not reduced to
polynomials.

\vskip0.2truein

{\bf 5. Dual SQCD}

As the main application of the reduction method, we consider here the
reduction of multi-
parameter theories appearing in connection with duality. As a particular
example, we discuss
the dual magnetic theory associated with SQCD \cite{SEI,APS}. While SQCD, as the `electric'
theory, has the
gauge coupling $g_e$ as the only coupling parameter, the dual `magnetic'
theory has two
parameters: the magnetic gauge coupling $g_m$ and a Yukawa coupling $\la_1$,
which measures
the strength of the interaction of color-singlet superfields with the
magnetic quark superfields.
It is our aim to discuss the reduced theories where the Yukawa coupling is
expressed in
terms of the gauge coupling.

For SQCD the gauge group is $SU(N_C)$ with N=1 supersymmetry. There are
$N_F$ quark superfields
${Q_i}$ and their antifields ${\tilde Q}^i,~i=1,2,...,N_F$ in the
fundamental representation.
For completeness and later reference, we give here the $\beta$-function
coefficients
for the electric SQCD theory:
\begin{eqnarray}
\beta_e (g_e^2)~=~\beta_{e0}~g_e^4~+~\beta_{e1}~g_e^6~+~ \cdots ,
\end{eqnarray}
with
\begin{eqnarray}
\beta_{e0}
~&=&~(16\pi^2)^{-1} (-3N_C ~+~ N_F ) \cr
\beta_{e1}~&=&~(16\pi^2)^{-2} \left( 2N_C(-3N_C + N_F) ~+~
4N_F \frac{N_C^2 - 1}{2N_C} \right) ~.
\end{eqnarray}

The corresponding dual magnetic theory is constructed mainly on the basis of
the anomaly
matching conditions \cite{SEI,SEN,IOS}. It involves the gauge group
$G^d=SU(N_C^d)$ with $N_C^d=N_F-N_C$.
Here $N_F$ is the number of quark superfields ${q_i},~{\tilde
q}^i,~i=1,2,...,N_F$
in the fundamental representation of $G^d$. Because both theories must have
the same global
symmetries, the number of flavors $N_F$ should be the same for SQCD and it's
dual.
As we have mentioned, duality requires a non-vanishing Yukawa coupling in
the form of
a superpotential
\begin{eqnarray}
W~=~\sqrt{\la_1} M^i_j q_i {\tilde q}^j.
\end{eqnarray}
The $N_F^2$ gauge singlet superfields $M^i_j$ are independent and cannot be
constructed
from $q$ and ${\tilde q}$. The superpotential not only provides for the
coupling of the
$M$ superfield, but also removes a global $U(1)$ symmetry acting on $M$,
which would have no
counterpart in the electric theory.

In the following, we will be dealing essentially only with the magnetic
theory. For convenience,
we therefore write $g$ in place of $g_m$ for the corresponding gauge
coupling. We also omit the subscript $m$ for the $\beta$-function
coefficients. Then the
$\beta$-function expansions of the magnetic theory are 
\begin{eqnarray}
\beta(g^2, \lambda_1)~&=&~\beta_{0}~g^4~+~(\beta_{1}~g^6~+~
\beta_{11}~g^4\lambda_1) ~+~\cdots \cr
\beta_{1} (g^2, \lambda_1)~&=&~c_1^{(0)} g^2 \lambda_1 ~+~
c_{11}^{(0)} \lambda_1^2 ~+~\cdots ~.
\end{eqnarray}
The coefficients are given by \cite{DSP,RDT,EGG}, \cite{AND}
\begin{eqnarray}
\beta_{0}~&=&~(16\pi^2)^{-1} (3N_C - 2N_F ) \cr
\beta_{1}~&=&~(16\pi^2)^{-2} \left( 2(N_F - N_C)(3N_C - 2N_F) ~+~
4N_F \frac{(N_F- N_C)^2 - 1}{2(N_F - N_C)} \right) \cr
\beta_{11}~&=&~(16\pi^2)^{-2} \left( -2N^2_F \right) \cr
c_1^{(0)}~&=&~(16\pi^2)^{-1} \left( -4 \frac{(N_F - N_C)^2 - 1}{2(N_F -
N_C)} \right) \cr
c_{11}^{(0)}~&=&~(16\pi^2)^{-1} \left(3N_F - N_C) \right) ~.
\end{eqnarray}

Already at the one-loop level, we see some important features
from Eqs.(37),(40). In the interval
\be
\frac{3}{2} N_C ~<~ N_F ~<~3N_C ,
\en
both theories are asymptotically free at large momenta, in particular the
magnetic theory for $N_F>\frac{3}{2}N_C$. For $N_F>3N_C$, the electric
theory
is not asymptotically free in the UV but in the IR, where the magnetic
version remains strongly coupled. Hence we expect that the original electric
excitations are present in the `physical' state space. The situation is
reversed
for $N_F<\frac{3}{2} N_C$, where the electric quanta are confined, and the
elementary
magnetic excitations describe the system, at least for $N_F>N_C+2$ where the
dual theory
exists which is the `free magnetic region'. This is the duality picture as
proposed by Seiberg,
with both theories describing the same physical system.

In the {\it conformal window} given in Eq.(41), the electric as well as the
magnetic theory
are in an interacting non-Abelian Coulomb phase, and it is indicated that
they both
have non-trivial conformal fixed points at zeroes of the exact
$\beta$-functions. At these fixed
points the theories are actually equivalent. Near an endpoint of the window,
in the
infrared limit, one theory may be in a weak coupling situation, and the
other, dual theory
in  a strong coupling regime. Since both theories represent the same system,
we can
describe the strongly coupled field theory by the weakly coupled dual. The
free excitations
of the latter may be considered as composites of those of the former theory.

Within the framework of this duality picture, the system undergoes an
important phase
transition at the point $N_F=\frac{3}{2}N_C$. As has already been mentioned
above, below this point the elementary electric
quanta are confined in the sense that they are not elements of the physical
state space. In the original electric theory, the transition at $N_F=\frac{3}{2}N_C$
is not apparent from the $\beta$-function coefficients, in contrast to
the phase change at $N_F = 3N_C$, where $\beta_{e0} = 0$. But in the duality
picture, we have $\beta_{0} = 0$ at $N_F=\frac{3}{2}N_C$ for the magnetic theory,
and this is the indication for the phase transition of the system.

Many years ago, we have obtained the phase transition of SQCD
at $N_F=\frac{3}{2}N_C$ by using
a rather different method \cite{LOP}. It involves analyticity and superconvergence of
the gauge field
propagator, as well as the BRST-cohomology in order to define the physical
state space
of the theory \cite{ROC,DSP}. The superconvergence 
relations, where they exist, are exact. They connect
long and short distance information, and are not valid in perturbation
theory \cite{WZS,ROS,OWG}.

The asymptotic form of the gauge field propagator
is governed by the ratio $\gamma_{00}/\beta_0$, where $\gamma_{00}$ is the
anomalous dimension of the gauge field (not the superfield) at the fixed
point $\alpha=0$.
Here $\alpha \geq 0$ is the conventional gauge parameter. Because this
parameter
is effectively a function of the momentum scale, it tends to a fixed point
asymptotically.
For example, in general covariant gauges, the discontinuity of the structure
function
has the asymptotic form 
\be
-k^2\rho(k^2,\kappa^2,g,\alpha)~&\simeq&~ C(g^2,\alpha)\left(-\beta_0
~ln\frac{k^2}{\kappa^2}\right)^{-\gamma_{00}/\beta_0},
\en
which is independent of $\alpha$ with the possible exception of the 
coefficient.

For the discussion of confinement using the BRST cohomology or the quark-
antiquark potential, it is most convenient to work in the Landau gauge,
where the superconvergence relation is of the form 
\be
\int_{-0}^{\infty} dk^2 \rho(k^2,\kappa^2,g,0)~=~0,
\en
provided $\gamma_{00}/\beta_0>0$. For general gauges $\alpha \geq 0$, 
the relation is the same except that the right hand side is given by
$\alpha/\alpha_0$, were $\alpha_0 = - \frac{\gamma_{00}}
{\gamma_{01}}$, with $\gamma_0(\alpha) = \gamma_{00} + \alpha \gamma_{01}$
\cite{OWG}. 

In contrast to the duality arguments, the superconvergence method is
applicable to
non-supersymmetric theories like QCD, where the interval corresponding to
the
window (53) for SQCD is given by
\be
\frac{13}{4} N_C ~<~ N_F ~<~\frac{22}{4}N_C ,
\en
For $N_F<\frac{13}{4} N_C$ for QCD and for $N_F<\frac{3}{2} N_C$ for SQCD,
our arguments show that the transverse gauge field excitations are not
elements of the
physical state space and hence confined. With some further arguments one can
extend
this result to quark fields.

For SQCD and similar theories, the connection between duality and
superconvergence
results is {\it quantitative}. 
For electric and magnetic SQCD, we have the anomalous dimensions
\be
\gamma_{e00}&=&(16\pi^2)^{-1}(-\frac{3}{2}N_C+N_F ) \cr
\gamma_{m00}&=&(16\pi^2)^{-1} \frac{1}{2}(-3N_C+N_F )~ ,
\en
and with the $\beta$-function coefficients from Eqs.(37),(40), we obtain
the relations \cite{NVS,DSP}
\be
\beta_{m0}(N_F)&=&-2\gamma_{e00}(N_F) \cr
\beta_{e0}(N_F)&=&-2\gamma_{m00}(N_F)~,
\en
where the argument $N_F$ on both sides refers to matter fields with
different quantum numbers corresponding to electric and magnetic gauge
groups. We have restored the subscript $m$ for these duality relations.
We see that $\gamma_{e00}(N_F)$ changes sign at the same point
$N_F=\frac{3}{2}N_C$ as $\beta_{m0}(N_F)$, and the ratio $\gamma_{e00}(N_F)
/\beta_{e0}(N_F)$ is positive below this point, indicating superconvergence
and confinement as discussed before.

The exact relations (46) are an indication, that the anomalous dimension
coefficients of the gauge fields at the fixed point $\alpha=0$ may have a more
fundamental significance, similar to the one-loop $\beta$-function coefficients.

Our discussion about the relation of superconvergence and duality results 
can be extended to similar supersymmetric gauge theories with other gauge
groups \cite{DSP,ROP,MOT}. The results are analogous. However, in the
presence of
matter superfields in the adjoint representation \cite{KUT},
the problem is more complicated. There the construction of dual theories requires
a superpotential already for the original electric theory, and a corresponding
reduction of couplings would be called for. Also the application of the
superconvergence arguments is not straight forward. 
These cases deserve further study. 

Duality in general superconformal theories has
been discussed in \cite{KLZ}, and for softly broken SQCD in \cite{KKK}.  

\vskip0.2truein

{\bf 6. Reduced Dual SQCD}

The magnetic theory dual to SQCD contains two parameters, the gauge
coupling $g$ and the Yukawa coupling $\la_1$. We now want to apply the reduction
method described in the previous sections and express the coupling parameter
$\la_1$ as a function of $g^2$. With Eq.(7) we write 
\be
\lambda_1(g^2) = g^2 f_1(g^2)~, ~\mbox{with}~~ f_1(g^2) = f^0 + 
\sum_{l=1}^{\infty}
\chi^{(l)} g^{2l} ~.
\en
The essential one-loop reduction equation is then
\be
\beta_{0} f^0 ~=~ \left( c_{11}^{(0)} f^0 ~+~
c_1^{(0)} \right) f^0 ~.
\en
There are two solutions:
\be
f^0 = f_{01} =
\frac{\beta_{0} - c_1^{(0)}}{c_{11}^{(0)}}~~\mbox{and} ~~f^0 = f_{00} = 0~,
\en
where $f_{01}$ is a function of $N_C$ and $N_F$, and is given by
\be
f_{01}(N_C, N_F)~=~\frac{N_C\left( N_F - N_C - {2}/{N_C} \right)}
{(N_F - N_C)(3N_F - N_C)} ~,
\en
using the explicit expressions (40) for the coefficients.
Here and in the following, we do not consider possible additional
terms which vanish exponentially or faster \cite{RON}.
The criteria for the unique definition of the coefficients $\chi^{(l)}$ in the 
expansion (47) are given by
\be
\left( M(f^0)- l \beta_{0}\right)
 \not = 0
   ~~ for ~~l=1,2,\ldots
\en  
with
\be
M (f^0) = c_{1}^{(0)} + 2 c_{11}^{(0)} f^0 -\beta_{0}~.
\en
Upon substitution of the solutions (49) and the explicit form of the 
coefficients from Eqs.(40), wie find
\be
M(f_{01}) - l\beta_{0} &=& -\beta_{0}(\xi + l)  \cr
M(f_{00}) - l\beta_{0} &=& +\beta_{0}(\xi - l)~,
\en
with $\beta_{0}$ from Eq.(40) and $\xi$ as a function of $N_C$ and $N_F$ given
by
\be
\xi(N_C, N_F)~=~\frac{N_C\left( N_F - N_C - {2}/{N_C} \right)}
{(N_F - N_C)(2N_F - 3N_C)} ~.
\en
The equations for the coefficients $\chi^{(l)}$ are of the general
form given in Eqs.(15).
For $l+1$ loops, they are simply
\be
\left(M (f^0) - l \beta_{0} \right)
\chi^{(l)} = \left(\beta_l (f^0) f^0 -
\beta^{(l)} (f^0)\right) + X^{(l)} ,
\en
where $l=1,2,\ldots ~$, and where $f^0$ is to be replaced by the solutions
$f_{01}$ or $f_{00}$ respectively.
The $\beta$-function coefficients are as in Eq.(11) with appropriate substitutions.

In the following , we consider characteristic intervals in $N_F$ separately,
and concentrate on the {\it conformal window}.
 
We have already discussed the window $\frac{3}{2} N_C ~<~ N_F ~<~3N_C $, where
both SQCD and dual SQCD are asymptotically free at small distances. Considering
first the solution $f_{01}(N_C,N_F)$ as given in Eq.(50), we see that it is positive
in the window, as is the function $\xi(N_C,N_F)$. Since also $\beta_{0}<0$, the coefficients
in Eq.(55) do not vanish. Consequently the expansion coefficients $\chi^{(l)}$ are uniquely
determined and can be removed by a regular reparametrization transformation. We are
left with the explicit solution
\be
\lambda_1 (g^2) ~=~ g^2 ~f_{01}(N_C, N_F)~,
\en
with $f_{01}$ given by Eq.(50). The $\beta$-functions of the reduced theory,
as defined by the solution (56), are now simply given by Eqs.(9) and (10)) with the
argument $f$ of the coefficient functions replaced by $f_{01}(N_C N_F)$ ,
so
that
they are constants:
\be
\beta (g^2) = \beta(g^2, g^2 f_{01}) = \sum^\infty_{l=0}
\beta_{l} (f_{01}) (g^2)^{l+2}  ~, ~~
\beta_{1} (g^2) = f_{01} \beta(g^2) ~.
\en
The second relation follows from the reduction equation (4) with Eq.(56).
The coefficient $\beta_{0}$ is as given in Eq.(40), and for $\beta_1 (f_{01})$
we obtain explicitly \cite{DSP,RDT}

\be
(16\pi^2)^2 \beta_{1}(f_{01}) ~=~   2(N_F - N_C)(3N_C-2N_F) ~+~
4N_F \frac{(N_F-N_C)^2-1}{2(N_F-N_C)}  \cr
- 4N_F^2 ~\frac{N_C (N_F-N_C-2/N_C)}
{2(N_F-N_C)(3N_F-N_C)} ~.
\en
These relations are used later in connection with the infrared fixed point
of dual SQCD in the conformal window near $N_F = \frac{3}{2}N_C$.
We must note here, that for the expansion (57), in addition to $\beta_0$,
the two-loop coefficient $\beta_{1}(f_{01})$ is {\it reparametrization invariant}.
This result follows because $f_{01}$ satisfies the reduction equation
(48).   

It remains to consider the second solution presented in
Eq.(49), with $f^0=f_{00}=0$. In this case the second 
expression in Eq.(53) is relevant for the determination 
of the  higher coefficients in the expansion of $f_1(g^2)$.
There could be a zero if $\xi(N_C,N_F)$ is a positive
integer in the window. Generally however, this is not the 
case (at least for $N_C < 16$), with the characteristic 
exception of $N_C=3,N_F=5$,
where $\xi(3,5)=2$ and the magnetic gauge group is 
$SU(2)$. Ignoring this case, we have again the situation
that all coefficients $\chi^{(l)}$ are determined and can be 
removed by regular reparametrization. Then the second 
power series solution of the reduction equations is given
by
\be
\la_1(g^2)~\equiv~0~,
\en
and leads to a theory without superpotential. As we have
discussed earlier, this situation is not acceptable for the 
dual magnetic theory. 

Returning to the exceptional case with the magnetic gauge 
group $SU(2)$, we find that,
after reparametrization, it leads to a solution of the form
\be
\la_1(g^2) = Ag^6 + \chi^{(3)}g^8 + \cdots~,
\en
where the coefficient $A$ is undetermined, and the higher
ones are fixed once $A$ is given. They vanish if $A=0$.
 We do not discuss this case here any further.

Finally, we briefly consider possible `general' solutions
of the reduction equations. It turns out that for dual SQCD
there are no such solutions which asymptotically approach the
relevant polynomial solution $\la_1(g^2)=g^2f_{01}$ given in
Eq.(56).  The only general solution we obtain is associated
with the excluded polynomial solution $\la_1(g^2)\equiv0$.
It is given by 
\be
\la_1(g^2) = A (g^2)^{1+\xi} + \cdots,
\en
where $A$ is again an undetermined parameter with properties
analogous to those discussed above for Eq.(60). As we have pointed
out, the exponent $\xi$, as given in Eq.(54), is positive and
generally non-integer in the limit. The only exception is for
$N_C=3,N_F=5$, in which case we are back to the exceptional
solution (60) discussed above.

We see that, within the set of solutions of the reduction equations
for magnetic SQCD, the power series $\la_1(g^2) = g^2 f_{01}$
is the unique choice for duality. Ignoring the isolated  $SU(2)$
case, the second power series solution $\la_1(g^2) \equiv 0$ is 
excluded. The general solution (61), which is associated with it,
leads to asymptotic expansions of Green's functions involving non-
integer powers. This is not consistent with a conventional,
renormalizable Lagrangian formulation. Since there are no general 
solutions approaching the power solution  $\la_1(g^2) = g^2 f_{01}$,
the latter is isolated or unstable.

From the one- and two-loop expressions for the $\beta$-functions
of the electric and the 
reduced magnetic theories given in Eqs.(37) and (40), we can 
obtain some information about non-trivial infrared fixed points
in the conformal window \cite{BZK,STR}. These expansions are useful as long
as the fixed points occur for values of $N_C, N_F$ near the 
appropriate endpoint of the window. We find \cite{RDT}
\be
\beta_m (g^{*2})  ~=~ 0  ~~~~~\mbox{for}~~~~~ \frac {g^{*2}}{16\pi^2} 
~=~ \frac{7}{3} ~\frac{N_F-\frac{3}{2} N_C}
{\frac{N_C^2}{4}-1} ~+~\cdots~,
\label{29}
\en
and
\be
\beta_e (g_e^{*2}) ~=~  0 ~~~~~\mbox{for}~~~~~ \frac {g_e^{*2}}{16\pi^2} 
~=~ \frac{3N_C-N_F}
{6(N_C^2-1)} ~+~\cdots~, 
\label{28}
\en
for sufficiently small and positive  values of $3N_C-N_F$ and $N_F-\frac{3}{2} N_C$
respectively. Larger values of $N_C$ may be needed in order to have a useful
approximation. Higher order terms have been calculated and may be found in \cite{EGG}.

With the reduced dual theory depending only upon the magnetic gauge coupling, it
is straightforward to obtain the critical exponent $\gamma_m=\gamma_m(N_C,N_F)$ near the
lower end of the window at $N_F=\frac{3}{2}N_C$ \cite{EGG}. This exponent is relevant for
describing the rate at which a given charge approaches the infrared fixed point.
With Eqs.(40), (58) and (62), the lowest order term is given by
\be
\gamma_m &=& \left(\frac{d\beta_m(g^2)}{dg^2}\right)_{g^2=g^{*2}}~=~ \frac{14}{3}~
\frac{(N_F - \frac{3}{2}N_C)^2}{\frac{N_C^2}{4} - 1}~  +~ \cdots~ ,
\en
where we have written $g$ in place of $g_m$ as before.
For the electric theory in the window near $N_F=3N_C$, the corresponding expression
is
\be
\gamma_e &=& \left(\frac{d\beta_e(g_e^2)}{dg_e^2}\right)_{g_e^2=g_e^{*2}}~=~ \frac{1}{6}~
\frac{(3N_C - N_F)^2}{N_C^2 - 1}~  +~ \cdots~ .
\en
In both cases we refer to \cite{EGG} for the next order.

In this report we consider mainly the reduction of dual magnetic SQCD in the
conformal window. A detailed discussion of the situation in the 
{\it free magnetic phase } $N_C + 2 \leq N_F < \frac{3}{2}N_C$ may be found in
\cite{RDT}. This interval is non-empty for $N_F>4$. The electric theory is
UV-free and the magnetic theory IR-free. At low energies, it is the latter which 
describes the spectrum. Because of the lack of UV-asymptotic freedom, one may
be concerned that the magnetic theory may not exist as a strictly local field
theory. However, it can be considered as a long distance limit of an appropriate
brane construction in superstring theory, which can also confirm 
duality \cite{KUS}. Except
for special cases involving again $SU(2)$ as the magnetic gauge group, the unique
power series solution (56) remains the appropriate choice also for this phase.
It correspond here to the approach to the {\it trivial infrared fixed point}.
Below $N_F=N_C+2$ there is no dual magnetic gauge theory, and the spectrum
should contain massless baryons and mesons associated with gauge invariant fields.

In the free electric phase for $N_F>3N_C$, the magnetic theory remains UV-free, 
and the results of the reduction method are the same as in the conformal window.

\vskip0.2truein

{\bf 7. Conclusions}

In the application to Duality, we see that the reduction method is most helpful
in bringing out characteristic features of theories with superpotentials. In the
case of the dual of SQCD, we get an essentially unique solution of the reduction
equations, which corresponds to a renormalizable Lagrangian theory with an 
asymptotic power series expansion in the remaining gauge coupling. This dual
magnetic theory is asymptotically free. 
It is UV-free in the conformal window and above,
and IR-free in the free magnetic region below the window.
In this latter region, it describes
the low energy excitations. These can be considered as 
composites of the free quanta of the 
electric theory, which is strongly coupled there.

As we have mentioned before, dual theories can be obtained as appropriate
limits of brane systems \cite{KUS}. In these brane constructions, duality corresponds
essentially to a reparametrization of the quantum moduli space of vacua
of a given brane structure. It is of interest
to find out how the reduction solutions are related to special
features of these constructions, in particular as far as the unique 
power solution (56) is concerned.  

\vskip0.2truein

Besides the use of the reduction method 
in connection with duality, which we have described
in this article, there are many other theoretical as well as 
phenomenological applications. 
Examples of applications in more phenomenological 
situations are discussed in this volume by J. Kubo \cite{JKR}.

\vskip0.2truein

Without detailed discussions, we mention here 
only a few applications:

\vskip0.3truein

* Construction of gauge theories with ``minimal'' coupling of
Yang-Mills and matter fields \cite{OSZ}.

* Proof of conformal invariance (finiteness)  for $N=1$ SUSY  gauge
theories with vanishing lowest order $\beta $-function
on the basis of one-loop information \cite{LPS,LUZ}.

* Reduction of the infinite number of coupling parameters appearing
in the light-cone quantization method \cite{WPL}.

* Reduction in an effective field theory formulation of quantum
gravity and in effective scalar field theory \cite{MAC}.

\noindent We see that the reduction method can be used also
within the framework of non-renormalizable theories, where the number
of couplings is infinite a priori.

* Applications of reduction to the standard model
(non-SUSY) give values for the top-quark mass which are too small,
indicating the need for more matter fields \cite{KSZ}.

* Gauge-Yukawa unifications within the framework of SUSY GUT's.
Successful calculations of top-quark and bottom-quark masses within
the framework of finite and non-finite theories \cite{KMO,KMZ,JKR}. 

* Reduction and soft symmetry breaking parameters. In softly broken
$N=1$ SUSY theories with gauge-Yukawa reduction, one finds all order
renormalization group invariant sum rules for soft scalar masses
\cite{DRT,KKZ,JKR}. There are interesting agreements with results
from superstring based models.

There are other problems where the reduction scheme is a helpful and often
an important tool \cite{SLC}.

\vskip2.0truein
\centerline{ACKNOWLEDGMENTS}
It was a pleasure to have participated in the Ringberg Symposium
in June 1998, where I have presented this talk.
I would like to thank Peter Freund, Einan Gardi, Jisuke Kubo,
Klaus Sibold, Wolfhart Zimmermann and George Zoupanos 
for very helpful discussions. I am grateful to Jisuke Kubo
for correspondence concerning our respective contributions
to the Festschrift.
Thanks are also due to Wolfhart Zimmermann, and the
theory group of the Max Planck Institut f\"ur Physik - Werner Heisenberg
Institut - for their kind hospitality in M\"unchen.
This work has been supported in part by the National Science Foundation,
grant PHY 9600697.

\newpage

\end{document}